\documentclass[letterpaper,twocolumn,showpacs,floatfix,groupaddress,twocolumns]{revtex4-1}
\usepackage[latin1]{inputenc}
\usepackage{bm}
\usepackage{multirow}
\usepackage{amssymb}
\usepackage{amsbsy}
\usepackage{amsmath,  amsthm, amsfonts, mathrsfs}
\usepackage{graphicx}
\usepackage{epsfig}
\usepackage{placeins}
\usepackage{float}
\usepackage{trfsigns}
\usepackage[usenames,dvipsnames]{color}
\usepackage[colorlinks,linkcolor=blue,citecolor=blue,urlcolor=blue,breaklinks=true]{hyperref}
\makeatletter

\begin{document} 

 \title{Integral Fluctuation Theorem for Microcanonical and Pure States}
 
 \author{Robin Heveling}
 \email{rheveling@uos.de}
 \affiliation{Department of Physics, University of Osnabr\"uck, D-49069 Osnabr\"uck, Germany}

 \author{Jiaozi Wang}
 \email{jiaowang@uos.de}
 \affiliation{Department of Physics, University of Osnabr\"uck, D-49069 Osnabr\"uck, Germany}

 \author{Jochen Gemmer}
 \email{jgemmer@uos.de}
 \affiliation{Department of Physics, University of Osnabr\"uck, D-49069 Osnabr\"uck, Germany}

\begin{abstract}
We present a derivation of the integral fluctuation theorem (IFT) for isolated quantum systems based on some natural assumptions on transition probabilities. Under these assumptions of ``stiffness'' and ``smoothness'' the IFT immediately follows for microcanonical and pure quantum states. 
We numerically check the IFT as well as the validity of our assumptions by analyzing two exemplary systems. We have been informed by T. Sagawa et al. that he and his co-workers found comparable numerical results and are preparing a corresponding
paper, which should be available on the same day as the present
text. We recommend reading their submission.
\end{abstract}
\maketitle

\section{Introduction}
\vspace*{-2px}
\noindent
The second law of thermodynamics states that in isolated systems the total entropy can only increase. However, the second law is merely of statistical nature, i.e. there may be exceedingly rare but possible processes in which the entropy does indeed decrease. As system sizes become smaller, violations to the second law become more prominent. These deviations are not random, but obey themselves rigid rules, which are often summarized under the name of fluctuation theorems. Fluctuation theorems formulate and, in some sense, generalize the second law of thermodynamics by relating the entropy production of processes, which may take the system arbitrarily far away from equilibrium, to properties of the equilibrated system in a quantitative manner (however, they make no statement about the system's route to equilibrium).
Just as for the second law, the underlying mechanisms which render these theorems valid or invalid are still under discussion.
In this work, we will show that the validity of the IFT for microcanonical and pure quantum states follows from natural assumptions on transition probabilities we call ``stiffness'' and ``smoothness'' \cite{stiff2018,stiff2020}. In essence, stiffness states that transition probabilities are largely independent of the initial energies. Furthermore, smoothness states that individual transition probabilities are close to the average transition probability in some respective energy interval.\\

\vspace*{-2px}
\noindent
In Sect. \ref{sctift}, we recapitulate the formulation of the IFT for a general system-bath setup. Thereafter, in Sect. \ref{sctcg}, we formulate a coarse-grained version of the IFT. In Sect. \ref{sectstiff}, we introduce the notion of stiffness and show the validity of the coarse-grained IFT follows from the assumption of stiffness. Following, in Sect. \ref{sectsmooth}, we introduce the notion of smoothness. It is presented how the assumption of smoothness connects the microscopic and the coarse-grained IFT. In Sect. \ref{numerics}, we substantiate our theoretical considerations by analyzing two specific numeric examples. In Sect. \ref{conc} follows a brief conclusion.

\section{Integral Fluctuation theorem for composite systems}
\label{sctift}
\noindent
The purpose of this preliminary section is to formulate the integral fluctuation theorem for a generic system-bath setup with total time-independent Hamiltonian
\begin{equation}
\label{hamiltonian}
H = H_{\text{sys}} + H_{\text{bath}} +H_{\text{int}}\,,
\end{equation}
where $H_{\text{sys}}$ is the system Hamiltonian and $H_{\text{bath}}$ is the bath Hamiltonian. System and bath are allowed to interact via an interaction term $H_{\text{int}}$. The composite system is initialized in a product state
\begin{equation}
\label{initial}
\rho(0) = \rho_\text{sys}(0) \otimes \rho_\text{bath}(0)\,,
\end{equation}
i.e. system and bath are initially uncorrelated and then brought into contact at $t=0$ via $H_{\text{int}}$. The time evolution operator $U(t)=\exp (- \mathrm{i} H t)$ propagates the composite system unitarily in time (for the remainder of this text we set $\hbar$ to unity).
We assume that initial system state and initial bath state are both diagonal in the eigenbases of their respective Hamiltonians, such that Eq. \eqref{initial} may be rewritten as
\begin{equation}
\rho(0) = \sum_{k,b} P^k_{\text{ini}} W^{b}_{\text{ini}} |k,b\rangle \langle k,b|\,,
\end{equation}
where $|k,b\rangle =  |\varepsilon_{\text{sys}}^k\rangle \otimes |\varepsilon_{\text{bath}}^b\rangle$ are products of eigenstates of system and bath Hamiltonians. The quantity $P^k_{\text{ini}}$ is the initial weight distribution over the energy eigenstates of the system and, respectively, $W^{b}_{\text{ini}}$ is the weight distribution over the energy eigenstates of the bath.\\

\noindent
A central operator of interest is the entropy production operator
\begin{equation}
\sigma(t) = - \log \rho_{\text{sys}}(t) + \beta H_{\text{bath}}\,,
\end{equation}
where $\beta=1/T$ is the inverse temperature of the bath ($k_\text{B}=1$) and $\rho_{\text{sys}}(t) = \text{Tr}_\text{bath}\{\rho(t)\}$ is the reduced density operator of the system at time $t$. The operator $\sigma(t)$ is explicitly time-dependent due to the first term. The eigenvalues of $\sigma(t)$ are given by
\begin{equation}
\label{evs}
\sigma^{j,a}(t) = - \log P^j_\text{sys}(t) + \beta \varepsilon_{\text{bath}}^a
\end{equation}
and the eigenstates by $|\sigma^{j,a}(t)\rangle$. We consider an ensemble average of differences in projective measurement outcomes at the initial time $t=0$ and some final time $t$ (denoted by double brackets $\langle\hspace*{-1.5px}\langle\, \bullet \,  \rangle\hspace*{-1.5px}\rangle$). Given some arbitrary but nicely behaved function $f$, the ensemble average of $f(\sigma(t))$ is defined by
\begin{align}
\label{ansembleav}
\langle\hspace*{-1.5px}\langle f(\Delta \sigma) \rangle\hspace*{-1.5px}\rangle
=\sum_{j,k,a,b} f(&\sigma^{j,a}(t)-\sigma^{k,b}(0)) \\ &\times P^k_{\text{ini}} W^{b}_{\text{ini}} R(jk,ab)\,,\nonumber
\end{align}
where $R(jk,ab;t)= |\langle \sigma^{j,a}(t) | U(t) |\sigma^{k,b}(0) \rangle|^2$ is the probability to transition from an initial state $|\sigma^{k,b}(0) \rangle$ to a final state $|\sigma^{j,a}(t) \rangle$. For brevity, we drop the explicit time dependence of the quantity in double brackets and in the argument of the transition probabilities.  For $f=\text{id}$,\;\; Eq. \eqref{ansembleav} yields the average entropy production $\langle\hspace*{-1.5px}\langle \Delta \sigma \rangle\hspace*{-1.5px}\rangle$, which can be written as a standard quantum mechanical expectation value
\begin{equation}
\langle\hspace*{-1.5px}\langle \Delta \sigma \rangle\hspace*{-1.5px}\rangle =\langle\sigma(t)-\sigma(0)\rangle = \Delta S_\text{sys}+ \beta \Delta U_\text{bath}\,,
\end{equation}
where the first term is the change in von Neumann entropy of the system and the second term the heat emitted from the bath. An important quantity related to the average entropy production is $\langle\hspace*{-1.5px}\langle e^{-\Delta\sigma}\rangle\hspace*{-1.5px}\rangle$, which is defined by Eq. \eqref{ansembleav} by setting $f(x)=\text{exp}(- x)$. This quantity (hereafter referred to as the IFT quantity) may be used to judge if the integral fluctuation theorem
\begin{equation}
\label{ift}
\langle\hspace*{-1.5px}\langle e^{-\Delta\sigma}\rangle\hspace*{-1.5px}\rangle = 1
\end{equation}
holds (if the initial energy distribution of the bath is canonical, Eq. \eqref{ift} is known to hold exactly). The IFT implies the second law of thermodynamics in the sense that the average entropy production (between some initial and some final point in time) is positive. This can be obtained by plugging Eq. \eqref{ift} into the Jensen inequality $e^{\langle\hspace*{-1.5px}\langle x\rangle\hspace*{-1.5px}\rangle} \leq \langle\hspace*{-1.5px}\langle e^{x}\rangle\hspace*{-1.5px}\rangle$ yielding  $\langle\hspace*{-1.5px}\langle \Delta \sigma \rangle\hspace*{-1.5px}\rangle \geq 0$. 
Applying the two point measurement scheme \cite{muka}  to the IFT quantity, i.e. averaging the exponentials of differences in measurement outcomes of $\sigma(t)$, yields
\begin{align}
\label{ift2}
\langle\hspace*{-1.5px}\langle e^{-\Delta\sigma}\rangle\hspace*{-1.5px}\rangle &= \sum_{j,k,a,b} \dfrac{P^j_{\text{fin}}}{\vphantom{\hat{f}}P^k_{\text{ini}}}  e^{-\beta (\varepsilon_{\text{bath}}^a-\varepsilon_{\text{bath}}^b)} P^k_{\text{ini}} W^{b}_{\text{ini}} R(jk,ab)\nonumber\\
&= \sum_{j,k,a,b} P^j_{\text{fin}} W^{b}_{\text{ini}}\, e^{-\beta (\varepsilon_{\text{bath}}^a-\varepsilon_{\text{bath}}^b)}R(jk,ab)\,,
\end{align}

\noindent
where $P_{\text{fin}}^j=P^j_\text{sys}(t)$. Importantly, even though the $P^k_{\text{ini}}$'s cancel out at the second equality sign, the sum over $k$ must still be kept (otherwise the IFT does not even hold for canonical initial states). For system-bath setups treatable with exact diagonalization techniques, Eq. \eqref{ift2} constitutes a convenient formula to check whether the IFT holds.


\newpage
\section{coarse-graining the Integral fluctuation theorem}
\label{sctcg}
\noindent
In this section, we will derive a ``coarse-grained'' version of the IFT similar to Eq. \eqref{ift2}. In general, the IFT addresses a time-dependent process described by some Hamiltonian $H(t)$. Here, we set the initial Hamiltonian $H(0)=H_\text{ini}$ and final Hamiltonian $H(T)=H_\text{fin}$ equal to the uncoupled Hamiltonian $H_{\text{unc}}=H_{\text{sys}} + H_{\text{bath}}$. The protocol then reads as follows, between the initial time $t=0$ and final time $t=T$ the interaction $H_{\text{int}}$ is instantaneously switched on, which induces transitions between the eigenstates of the uncoupled Hamiltonian. In the following, we drop the subscripts ``ini'' and ``fin'' since initial and final Hamiltonian coincide. The eigenvalue equations for the initial and the final Hamiltonian read
\begin{align}
H_{\text{unc}} |k,b \rangle = \varepsilon^{k,b} |k,b \rangle
\end{align}
with eigenvalues $ \varepsilon^{k,b}= \varepsilon_{\text{sys}}^k + \varepsilon_{\text{bath}}^b$. The eigenstates are $|k,b \rangle=|\varepsilon_{\text{sys}}^k\rangle\otimes |\varepsilon_{\text{bath}}^b\rangle$.
The density of states (DOS) of the bath is given by
\begin{align}
\Omega_\text{bath}(E_\text{bath})= \sum_{b}  \delta(E_\text{bath}-\varepsilon^{b}_\text{bath}) \,.
\end{align}
This description will now be extended to finite energy resolutions. We assume that the system only comprises a few energy levels, e.g. a single spin. Therefore, we will resort to dividing just the energy scale of the bath into bins of finite size. To this end, we will introduce the bin size $\Delta$ (not to be confused with the delta used in the notation for the ensemble average) and divide the energy scale of the bath into intervals according to $E^B_{\text{bath}} = [(B-1/2) \Delta, (B+1/2) \Delta]$, where $B=...,-2,-1,0,1,2,...$\,\,. The midpoint of the energy interval $E^B_{\text{bath}}$ of width $\Delta$ is $B\Delta$. The bin size $\Delta$ should be small compared to the energy scale of the bath, but large compared to its level spacing. As of yet, individual eigenvalues were denoted by an ``$\varepsilon$''. Now, energy intervals are denoted by an ``E'' and enumerated by capitalized indices. \\



\noindent
The probability $R(jk,Ab)$ to transition from an initial eigenstate $|k,b\rangle$ to any state $|j,a\rangle$ with $\varepsilon^a_\text{bath} \in E^A_\text{bath}$, i.e. to a range of bath eigenstates, is obtained by
\begin{equation}
\label{RjkAb}
R(jk,Ab)=\sum_{a:\,\varepsilon_\text{bath}^{a} \in E_\text{bath}^{A}} R(jk,ab)\,.
\end{equation}
In a similar fashion, we define the \textit{average} probability to transition from an initial state $|k,b\rangle$ with $\varepsilon^b_\text{bath} \in E^B_\text{bath}$ to a final energy interval $E^{A}_\text{bath}$ by 
\begin{align}
\overline{\vphantom{|}R}(jk,AB)= \dfrac{1}{\Omega^{B}_{\text{bath}}}
\sum_{b:\,\varepsilon_\text{bath}^{b} \in E_\text{bath}^{B} } R(jk,Ab)\,,
\end{align}
\,
\noindent
where
\begin{equation}
\Omega^{B}_{\text{bath}} = \int_{(B-1/2)\Delta}^{(B+1/2)\Delta} \Omega_\text{bath}(E_\text{bath}) \,\text{d}E_\text{bath}
\end{equation}
is the number of bath energy eigenvalues $\varepsilon^{b}_\text{bath}$ within the energy interval $E^{B}_\text{bath}$. The coarse-grained transition probabilities satisfy
\begin{equation}
\label{trans}
\sum_{j,A} \overline{\vphantom{|}R}(j k,A  B) = 1\,.
\end{equation}
In addition, we need the concept of microreversibility, which will serve as starting point for our derivation. In short, a certain setup is microreversible, if the relevant observables and the Hamiltonian are real in the working basis. In the case of microreversibility, considerations along the lines of microcanonical fluctuation theorems yield
\begin{equation}
\label{mft2}
\dfrac{\overline{\vphantom{|}R}(j  k,A  B)}{\vphantom{|}\widetilde{R}(j  k,A  B)}=\dfrac{\Omega^{A}_{\text{bath}}}{\Omega^{B}_{\text{bath}}}\,,
\end{equation}
where the tilde indicates the average probability of a transition of a time reversed process described by the time evolution operator $\widetilde{U}(t)$ of a backwards protocol implemented by $\widetilde{H}=H(T-t)$. In the case at hand we have that $\widetilde{U}(t) = U(t)$. 
We start our derivation by assuming microreversibility, i.e. that Eq. \eqref{mft2} is fulfilled. Additionally assuming an exponentially growing density of states of the bath we arrive at
\begin{equation}
\label{start}
\dfrac{\overline{\vphantom{|}R}(j  k,A  B)}{\widetilde{R}(j  k,A  B)}=\dfrac{e^{\beta \Delta A}}{\vphantom{\tilde{f}}e^{\beta \Delta B}}\,,
\end{equation}
where 
\begin{equation}
\label{back}
\widetilde{R}(j k,A  B)=\overline{\vphantom{|}R}(k  j,B  A)
\end{equation}
are the transition probabilities of the backwards motion. This relation only holds, if, in addition to microreversibility, the protocol is symmetric in time, which is the case here. \\

\noindent
By algebraic manipulation of Eq. \eqref{start}, multiplying with $P^j_{\text{fin}}W^B_{\text{ini}} $ and summing over $j,k,A,B$ we get that
\begin{align}
\label{iftint}
\sum_{j,k,A,B}& P_{\text{fin}}^j W_{\text{ini}}^B e^{-\beta \Delta (A-B)} \overline{\vphantom{|}R}(j  k,A  B)\\
=&\sum_{j,k,A,B} P^j_{\text{fin}}W^B_{\text{ini}} \widetilde{R}(j  k,A  B)\nonumber\,.
\end{align}
The l.h.s. of Eq. \eqref{iftint} looks similar to the r.h.s. of  Eq. \eqref{ift2}, such that we define the coarse-grained (``c.g.'') version of the IFT as
\begin{align}
\label{iftcg}
\hspace*{-15px}\langle\hspace*{-1.5px}\langle e^{-\Delta\sigma}\rangle\hspace*{-1.5px}\rangle_{\text{c.g.}} =\sum_{j,k,A,B}P_{\text{fin}}^j W_{\text{ini}}^B e^{-\beta \Delta (A-B)} \overline{\vphantom{|}R}(j  k,A  B)\,.
\end{align}
\newpage
\noindent
Indeed, Eq. \eqref{iftcg} may be interpreted as a ``coarse-grained" or ``microcanonical" version of $\langle\hspace*{-1.5px}\langle e^{-\Delta\sigma}\rangle\hspace*{-1.5px}\rangle$ in the sense that: \textit{i}. energy changes in the bath are  now  counted on the level of energy intervals rather than individual energy eigenvalues, \textit{ii}. the transition probabilities now apply to transitions of microcanonical initial states restricted to the initial interval $E^B_\text{bath}$ to the final energy interval $E^A_\text{bath}$ rather than to transitions between individual eigenstates and \textit{iii}. the initial probabilities for the bath are now probabilities to find the bath in the respective energy interval rather than in the corresponding eigenstate. One may be inclined to think that Eq. \eqref{iftint} in general becomes Eq. \eqref{ift2} in the limit of small energy intervals. But this sentiment is flawed, since Eq. \eqref{mft2} and thus Eq. \eqref{iftint} rely on notions that ultimately break down in the limit of small energy intervals. \\

\section{Validity of the coarse-grained integral fluctuation theorem via stiffness}
\label{sectstiff}
\noindent
In this section we will define the property of stiffness for coarse-grained transition probabilities, which we will utilize to show that $\langle\hspace*{-1.5px}\langle e^{-\Delta\sigma}\rangle\hspace*{-1.5px}\rangle_{\text{c.g.}}=1$ holds for all possible $W^B_\text{ini}$.
We call a transition probability from an initial energy interval $E^{B}_\text{bath}$ to a final energy interval $ E^{A}_\text{bath}$ \textit{stiff} (for  given $j,k$), if we have that
\begin{equation}
\label{stiff}
\overline{\vphantom{|}R}(jk,AB) = \overline{\vphantom{|}R}(jk,A-B)\,,
\end{equation}
i.e. the probability to transition from the initial energy interval $E^{B}_\text{bath}$ to a state within the final energy interval $E^{A}_\text{bath}$ is only a function of the difference in energies. 
The above definition of stiffness implies that
\begin{equation}
\label{stiffprop}
\sum_A \overline{\vphantom{|}R}(j k,A  B) = \sum_B \overline{\vphantom{|}R}(j  k,A  B)\,.
\end{equation}

\noindent
Now, plugging the occupation probability of an initial microcanonical bath state $W_{\text{ini}}^B=\delta_{B,B'}$ (which, in this case, completely ``fills'' one energy interval) into the r.h.s. of Eq. \eqref{iftint} yields
\begin{align}
\langle\hspace*{-1.5px}\langle e^{-\Delta\sigma}\rangle\hspace*{-1.5px}\rangle_{\text{c.g.}} &= \sum_{j,k,A,B} P^j_{\text{fin}}\delta_{B,B'} \widetilde{R}(j  k,A  B) \\
&=\sum_{j,k,A} P^j_{\text{fin}} \widetilde{R}(j  k,A  B')\nonumber\\
&= \sum_{j,k,A} P^j_{\text{fin}} \overline{\vphantom{|}R}(k  j,B' A)\nonumber\\\nonumber
&= \sum_{j,k,B'} P^j_{\text{fin}} \overline{\vphantom{|}R}(k  j,B'  A)\nonumber\\
&= \sum_j P^j_{\text{fin}} \sum_{k,B'} \overline{\vphantom{|}R}(k  j,B'  A)\nonumber\\
&= \sum_j P^j_{\text{fin}} = 1\nonumber\,.
\end{align}
To go to the second line the sum over $B$ was evaluated, at the third equality sign Eq. \eqref{back} was used, at the fourth equality sign the stiffness property, i.e. Eq. \eqref{stiffprop}, was applied, then the sums were factorized and Eq. \eqref{trans} was employed to reach the last line.
Thus, the coarse-grained IFT is shown for microcanonical initial states under the assumption of stiffness.\\

\noindent
Note that we started from Eq. \eqref{mft2}, which is known to be true for microreversible setups. Even if microreversibility is broken, e.g. by magnetic fields, Eq. \eqref{mft2} may still hold and serve as a starting point for our derivation. We conjecture that, even if Eq. \eqref{mft2} is violated, the IFT will still be fulfilled in systems featuring stiffness. In that sense, Eq. \eqref{mft2} is a convenient tool to derive the desired result, but not the essential ingredient.\\

\section{link between coarse-grained and microscopic integral fluctuation theorem via smoothness}
\label{sectsmooth}
\noindent 
In this section we will define the property of smoothness of transition probabilities and show that under the assumption of smoothness, Eq. \eqref{iftcg} actually becomes an arbitrarily good approximation of Eq. \eqref{ift2}. This holds even if the bath initially only occupies a single energy eigenstate.
In general, the probability to transition from some initial energy eigenstate to some final energy interval of course differs from the average transition probability to go from the initial energy interval (in which the initial energy lies) to the final energy interval. We denote the difference of these two quantities by $r$ according to
\begin{align}
\label{smoothie}
R(jk,Ab)= \overline{\vphantom{|}R}(jk,AB) +r(jk,Ab) \,.
\end{align}
\noindent
We call a set of transition probabilities from an initial state $|k,b\rangle$ to a final energy interval $E^{A}_\text{bath}$ \textit{smooth} (for given $j,k$), if we have that
\begin{equation}
\label{smooth}
r(j  k,A b) \approx 0
\end{equation}
for all $\varepsilon^{b}_\text{bath} \in E^{B}_\text{bath}$, i.e. all transition probabilities to go from a state with initial energy $\varepsilon^{b}_\text{bath}$ within the initial interval $E^{B}_\text{bath}$ to the final energy interval $E^{A}_\text{bath}$ are close to the average value of transition probabilities within that initial energy interval. \\

\noindent
To continue we need two more ingredients. Firstly, we will use that the exponential factor $\exp(-\beta (\varepsilon_{\text{bath}}^a-\varepsilon_{\text{bath}}^b))$ only changes negligibly when replacing the distance between individual energy eigenvalues with the distance between energy intervals, namely
\begin{align}
\label{exp}
e^{-\beta (\varepsilon_{\text{bath}}^a-\varepsilon_{\text{bath}}^b)} \approx e^{-\beta \Delta (A-B)}\,.
\end{align}
This can be achieved by making the bin size $\Delta$ sufficiently small or by increasing the temperature. Secondly, we use that the initial weight in an energy interval of the bath  $W^{B}_{\text{ini}}$ is obtained by summing all individual initial weights of eigenstates with energies in that initial energy interval, i.e.
\begin{align}
\label{wb}
 \sum_{b:\,\varepsilon_\text{bath}^{b} \in E_\text{bath}^{B}}W^{b}_{\text{ini}} = W^B_\text{ini}\,.
\end{align}
\noindent
The starting point is the r.h.s. of Eq. \eqref{ift2}. In the derivation below, firstly, the sums over $a$ and $b$ are split according to the belonging of $\varepsilon^a_\text{bath}$ and $\varepsilon^b_\text{bath}$ to their respective energy intervals $E^A_\text{bath}$ and $E^B_\text{bath}$. 
Then, Eq. \eqref{exp} is employed such that the exponential factor can be pulled to the front. 
Next, we recognize the sum over $a$ from Eq. \eqref{RjkAb}. To go to the next line, we plug in Eq. \eqref{smoothie} and abbreviate all terms linear in $r$ as $\mathcal{O}(r)$. Then, Eq. \eqref{wb} was used. Finally, we employ the assumption of smoothness, i.e. Eq. \eqref{smooth}, which allows to neglect terms linear in $r$.\\

\noindent
Thus, it can be seen that the two versions of the IFT indeed coincide if the transition probabilities are sufficiently smooth. To repeat, the initial energy interval can not be arbitrarily small, such that the  IFT for single energy eigenstates does not follow immediately. However, the validity of the IFT follows under the assumption of smoothness.\\
\begin{widetext}
\begin{align}
\label{intervalegal}
\langle\hspace*{-1.5px}\langle e^{-\Delta\sigma}\rangle\hspace*{-1.5px}\rangle
=&\sum_{j,k,a,b} P^j_{\text{fin}} W^{b}_{\text{ini}}\, e^{-\beta (\varepsilon_{\text{bath}}^a-\varepsilon_{\text{bath}}^b)} R(j  k,a  b)\\\nonumber
= &\sum_{j,k,A,B} \sum_{a:\,\varepsilon_\text{bath}^{a} \in E_\text{bath}^{A}} \sum_{b:\,\varepsilon_\text{bath}^{b} \in E_\text{bath}^{B}}P^j_{\text{fin}} W^{b}_{\text{ini}}\, e^{-\beta (\varepsilon_{\text{bath}}^a-\varepsilon_{\text{bath}}^b)} R(j  k,a  b)\\\nonumber
\approx &\sum_{j,k,A,B} P^j_{\text{fin}} e^{-\beta \Delta (A-B)} \sum_{b:\,\varepsilon_\text{bath}^{b} \in E_\text{bath}^{B}} W^{b}_{\text{ini}} \sum_{a:\,\varepsilon_\text{bath}^{a} \in E_\text{bath}^{A}}  R(j  k,a  b)\\\nonumber
= &\sum_{j,k,A,B} P^j_{\text{fin}}  e^{-\beta \Delta (A-B)} \sum_{b:\,\varepsilon_\text{bath}^{b} \in E_\text{bath}^{B}} W^{b}_{\text{ini}}\, R(j  k,A  b)\\\nonumber
=&\;\,\mathcal{O}(r) + \sum_{j,k,A,B} P^j_{\text{fin}}  e^{-\beta \Delta (A-B)} \sum_{b:\,\varepsilon_\text{bath}^{b} \in E_\text{bath}^{B}} W^{b}_{\text{ini}}\, \overline{R}(j  k,A  B) \\\nonumber
=&\;\,\mathcal{O}(r)+\sum_{j,k,A,B} P^j_{\text{fin}} W^{B}_{\text{ini}} e^{-\beta \Delta(A-B)} \overline{R}(j  k,A  B)\nonumber
\stackrel{r \rightarrow 0}{\longrightarrow} \langle\hspace*{-1.5px}\langle e^{-\Delta\sigma}\rangle\hspace*{-1.5px}\rangle_{\text{c.g.}}\nonumber
 \end{align}
 \end{widetext}

\section{Numerical Verification of the IFT}
\label{numerics}
\noindent
\noindent
In this section, we will present a numerical analysis on two exemplary systems, a hardcore boson model and and a transverse Ising model with defects. We will numerically check whether the IFT holds for both systems. Furthermore, we will investigate to what extent stiffness and smoothness are fulfilled/violated.\\

\noindent
As mentioned in Sect. \ref{sctift}, the initial state of the composite system will be a product state of a system state and an energy eigenstate of the bath corresponding to some inverse temperature $\beta$. In the following, we employ a microcanonical definition of temperature. For some given energy $E_0$, the inverse temperature will be determined by the exponent of an exponential fit of the DOS in the direct vicinity of $E_0$.\\

\noindent
The first setup of interest is a hardcore boson model, which was also the subject in Ref. \cite{yoda}. The system just consists of a single site and the bath is a quadratically ($4\times 4$) shaped lattice, yielding a total of $17$ sites. The system interacts with one corner of the bath. The relevant Hamiltonians read
\begin{equation}
H_\text{sys} = \omega n_0\quad\quad H_\text{int} = - \gamma' (c_0^\dagger c_1+c_1^\dagger c_0) 
\end{equation}
\begin{equation*}
H_\text{bath} = \omega \sum_i n_i - \gamma \sum_{\langle i,j\rangle}(c_i^\dagger c_j+c_j^\dagger c_i) +g \sum_{\langle i,j\rangle} n_i n_j\,,
\end{equation*}
where $c_j$ ($c_j^\dagger$) are annihilation (creation) operators of a boson on site $j$ and $n_j=c_j^\dagger c_j$ is the occupation number on site $j$. The double sums $\langle i,j\rangle$ run over horizontally and vertically neighboring sites with open boundaries. Note that, due to symmetry of the interaction term, the eigenstates of the entropy production operator $|\sigma^{k,b}(t)\rangle$ are equal to the eigenstates $|k,b\rangle$ of the uncoupled Hamiltonian. The parameters take values $\gamma=1.0$, $g=0.1$ and $\omega = 10.0$, while the interaction strength $\gamma'$ is varied. The initial bath energy is $\varepsilon_\text{bath}^{b_0}=29.286$ corresponding to an inverse temperature of $\beta=0.108$. The composite system is initialized in the product state
\begin{equation}
\label{ini}
\rho(0) = |1,b_0\rangle\langle 1,b_0| \,,
\end{equation}
i.e. initially the system site is occupied with probability one. Following Ref. \cite{yoda}, we restrict the dynamics to a specific particle sector, here $N=4$. Since the system site is initially occupied with probability one, there can only be three  particles left in the bath, i.e. $\sum_i\langle b_0|n_i|b_0\rangle = 3$, where the sum runs over all bath sites. The onsite potential is the same on every site, thus, it does not affect the occupation dynamics on any site.\\
\begin{figure}[b]
  \includegraphics[width=\linewidth]{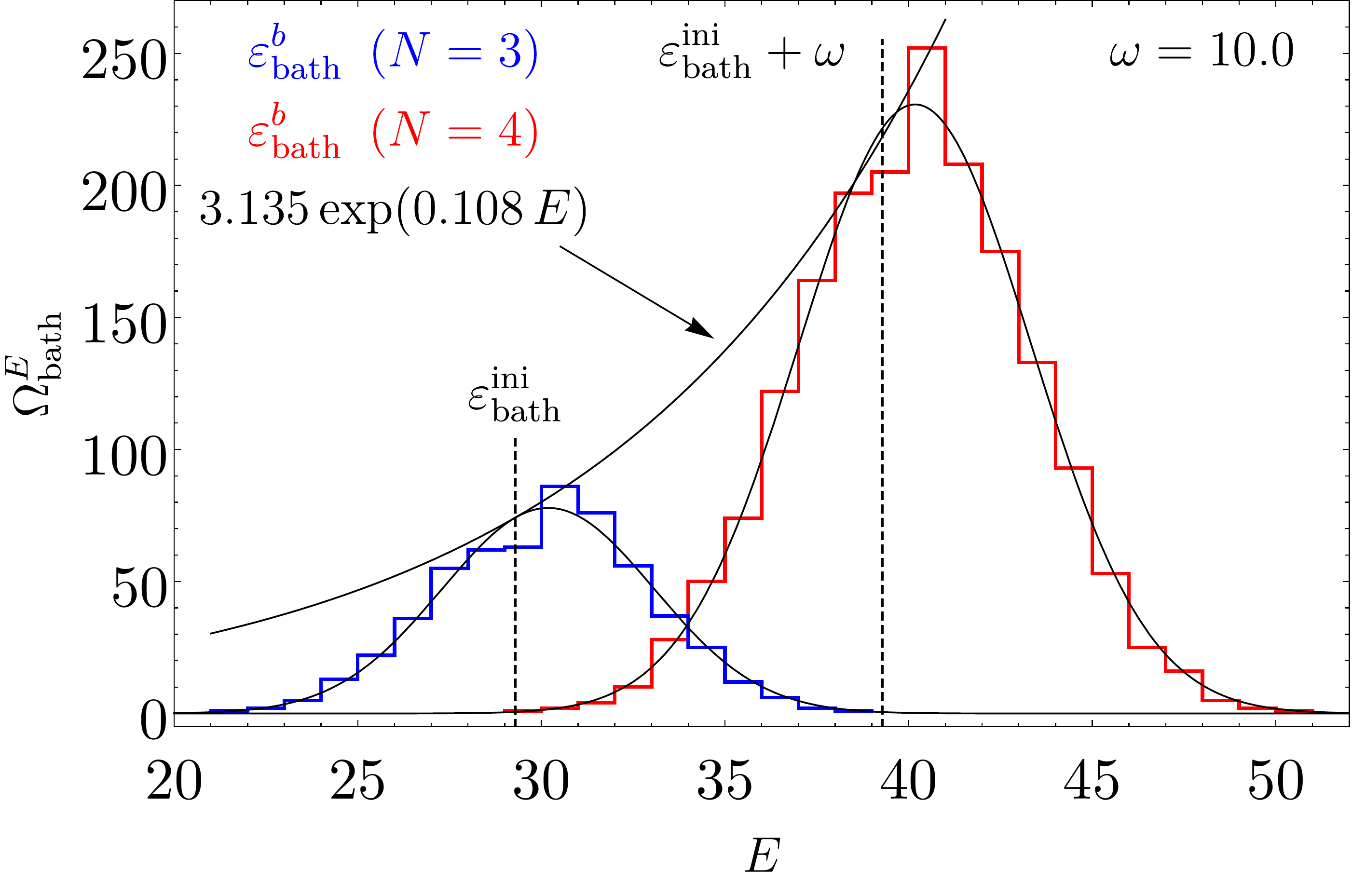}
  \caption{Histograms of bath energy eigenvalues in the $N=3$ and $N=4$ particle sectors fitted with Gaussians. The quantity $\Omega_\text{bath}^E$ indicates the number of eigenvalues within a bin of size $\Delta = 1.0$ at the energy $E$. The energetically accessible parts are well approximated by a single exponential function with exponent $\beta E$.}
  \label{pic34}
\end{figure}
\newpage
\begin{figure}[t]
  \includegraphics[width=\linewidth]{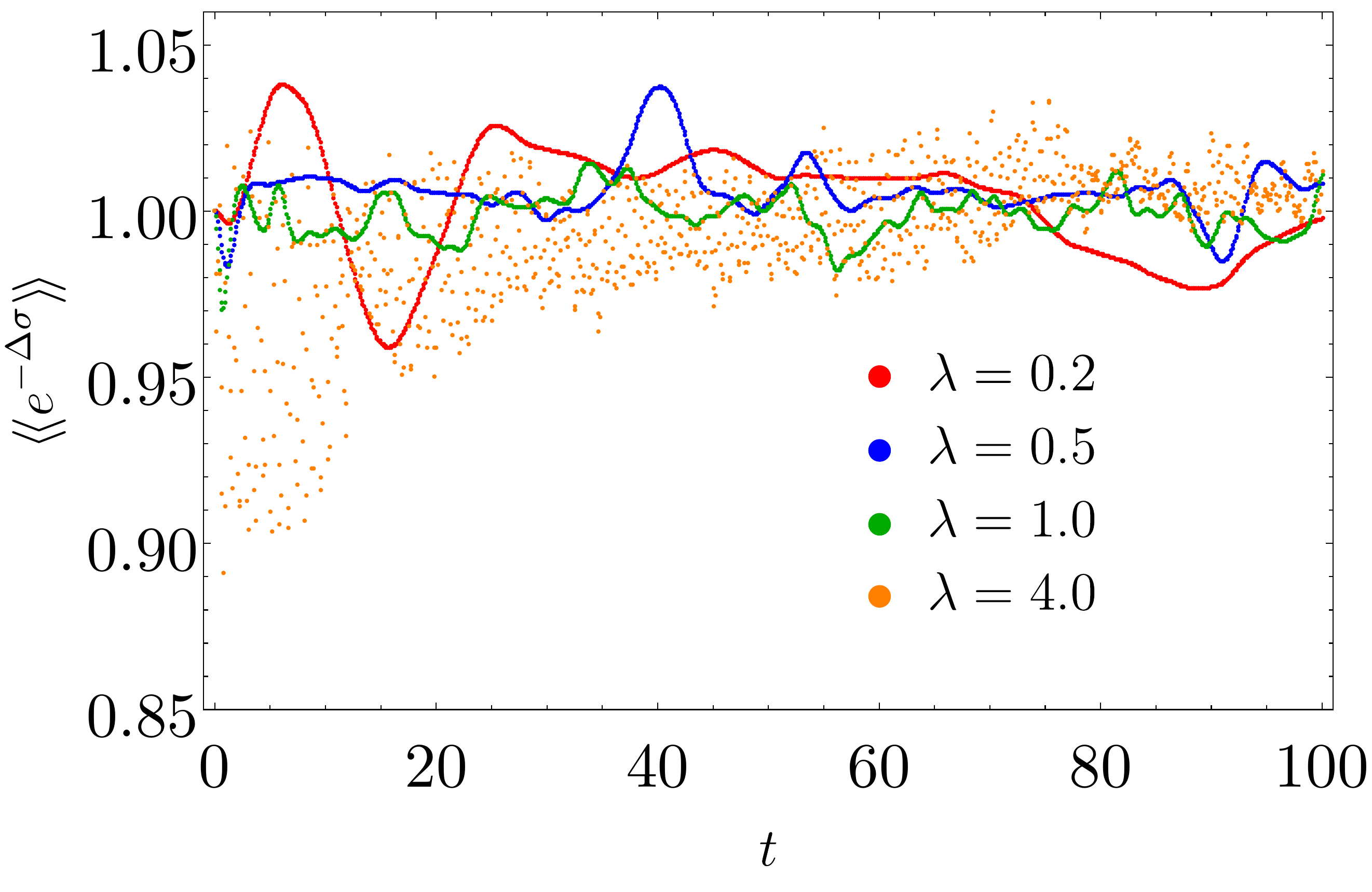}
  \caption{The IFT quantity $\langle\hspace*{-1.5px}\langle e^{-\Delta\sigma}\rangle\hspace*{-1.5px}\rangle$ plotted over time $t$ for the hardcore boson model. There are small deviations of about $0.05$ from the desired value of one. For the strongest interaction strength $\gamma'=4.0$ the initial deviations are larger.}
  \label{hbmodel}
\end{figure}
\noindent
However, it changes the entropy production. For $\omega=10.0$  the energetically relevant parts of the spectra of the $N=3$ and $N=4$ particle sectors can both be well approximated by a single exponential function, cf. Fig. \ref{pic34}. This property was crucial for our derivation of the coarse-grained IFT, cf. Eq. \eqref{start}. (Note that the $N=3$ particle sector is relevant due to the sum over $k$ in Eq. \eqref{ift2}).\\
\noindent
The temporal behavior of the IFT quantity is depicted in Fig. \ref{hbmodel} for various interaction strengths. The IFT quantity moderately fluctuates around the desired value of one with deviations of about $0.05$. For the strongest interaction $\gamma'=4.0$ the initial deviations are more pronounced. The  temporal behavior of the coarse-grained IFT quantity is shown in Fig. \ref{hbmodelcg} for the same cases. After some mild initial fluctuations, the deviations from one are smaller than in Fig. \ref{hbmodel}, suggesting that the property of smoothness is violated to some extent.
\begin{figure}[b]
  \includegraphics[width=\linewidth]{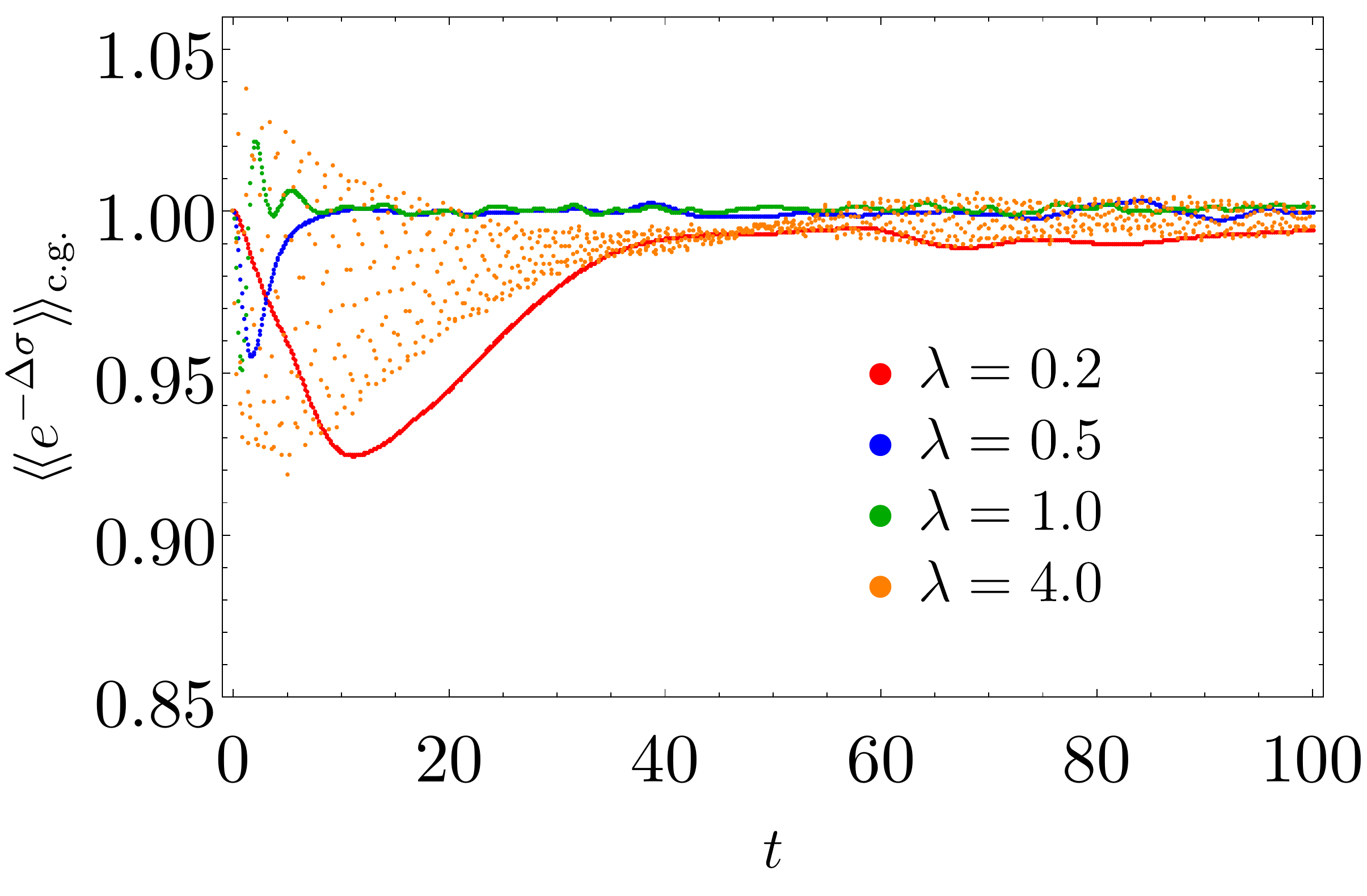}
  \caption{The coarse-grained IFT quantity $\langle\hspace*{-1.5px}\langle e^{-\Delta\sigma}\rangle\hspace*{-1.5px}\rangle_{\text{c.g.}}$ plotted over time $t$ for the hardcore boson model. The initial bath state is microcanonical, comprising $50$ eigenstates neighboring $\varepsilon_\text{bath}^{b_0}$. Deviations from one are visibly smaller than in Fig. \ref{hbmodel}.}
  \label{hbmodelcg}
  \end{figure}
  
\begin{figure}[t]
  \includegraphics[width=\linewidth]{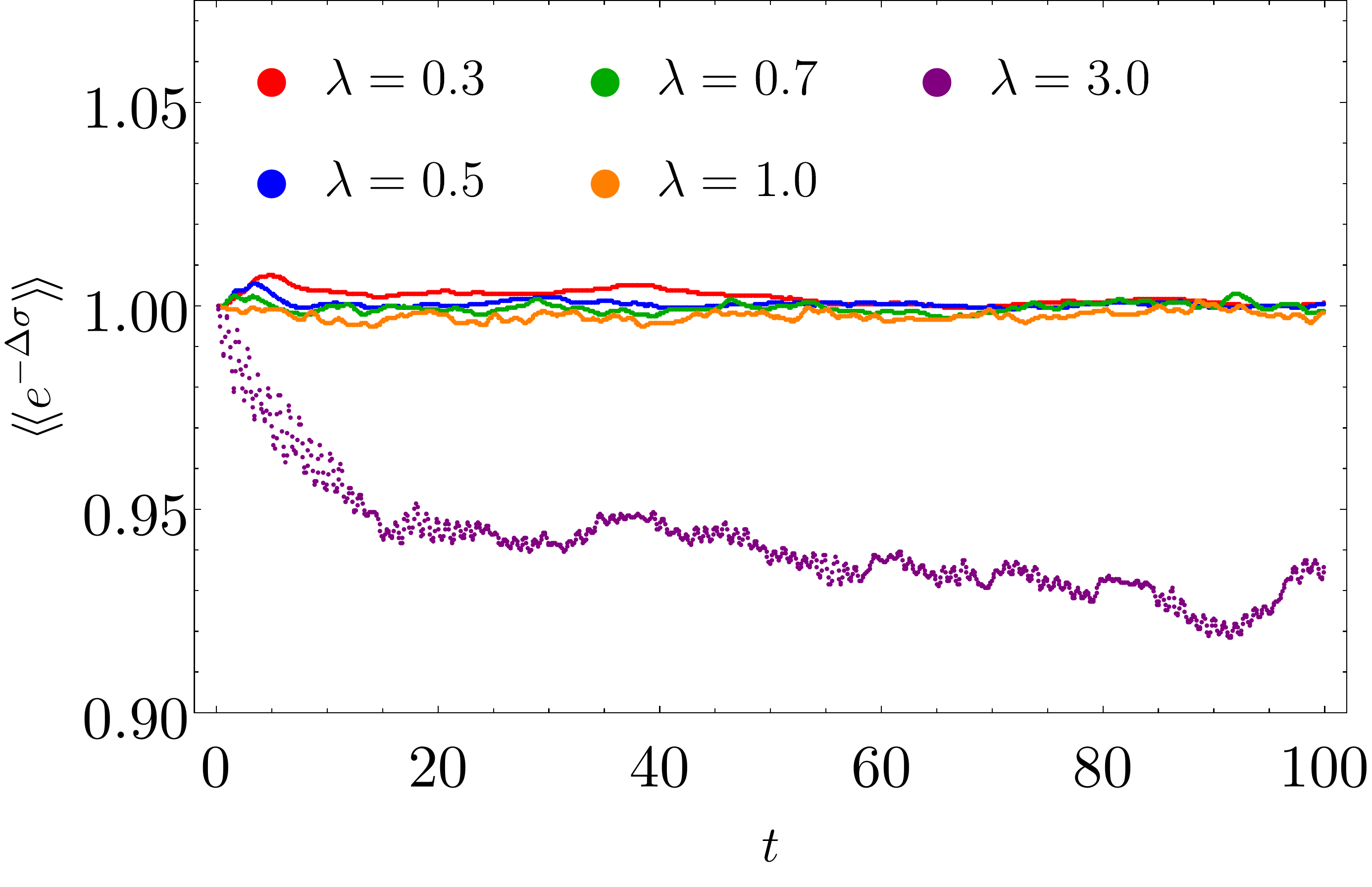}
  \caption{The IFT quantity $\langle\hspace*{-1.5px}\langle e^{-\Delta\sigma}\rangle\hspace*{-1.5px}\rangle$  plotted as a function of time for the Ising model. For weak to moderate interaction strengths the IFT is practically fulfilled.}
  \label{isingift}
\end{figure}

\begin{figure}[b]
  \includegraphics[width=\linewidth]{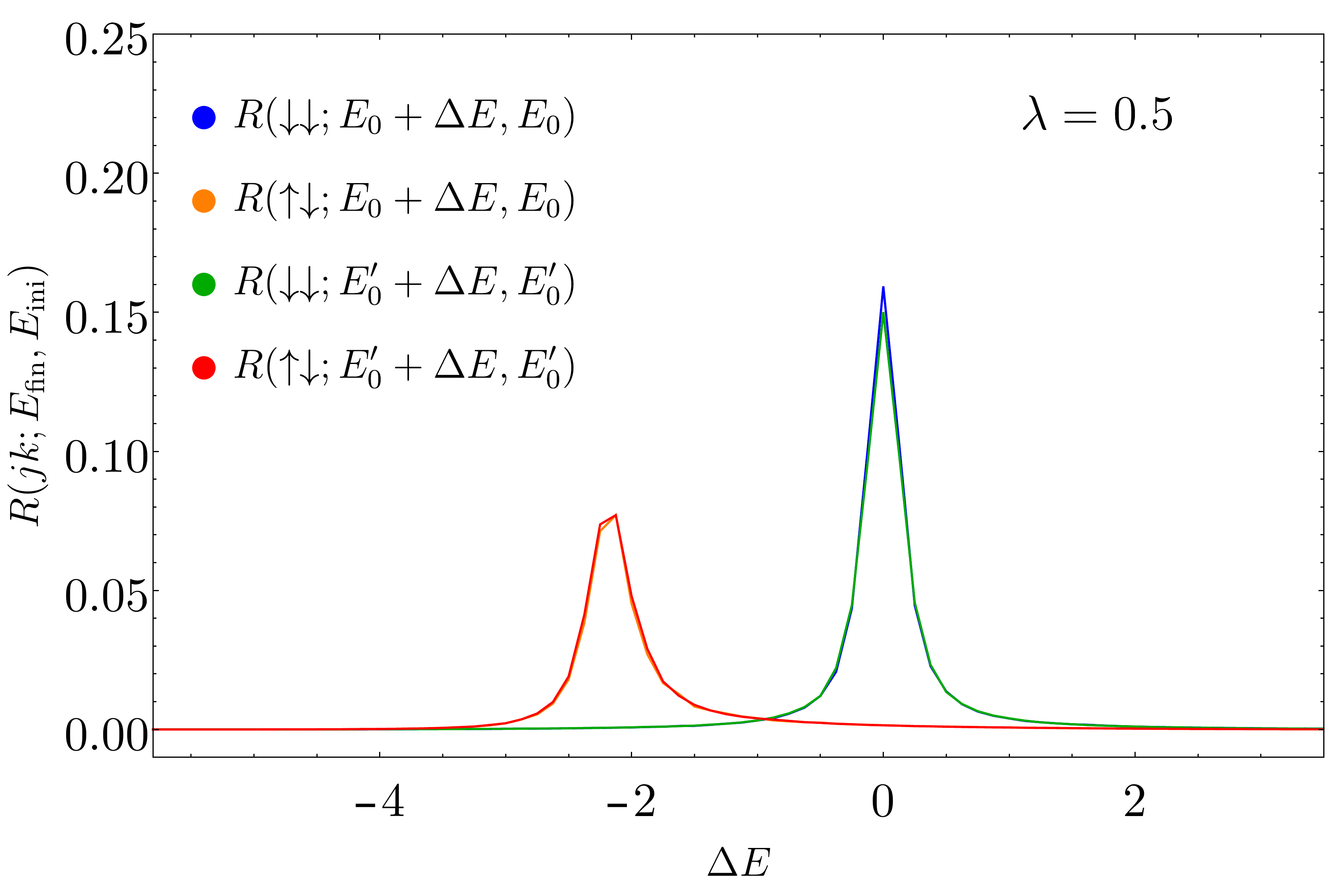}
  \caption{Transition probabilities for $\lambda=0.5$, $j=\,\uparrow,\downarrow$ and initial energy intervals $E_0,E_0'$. The transition probabilities coincide for the two different initial energies, which indicates the existence of stiffness.}
  \label{stiff5}
\end{figure}

\begin{figure}[t]
  \includegraphics[width=\linewidth]{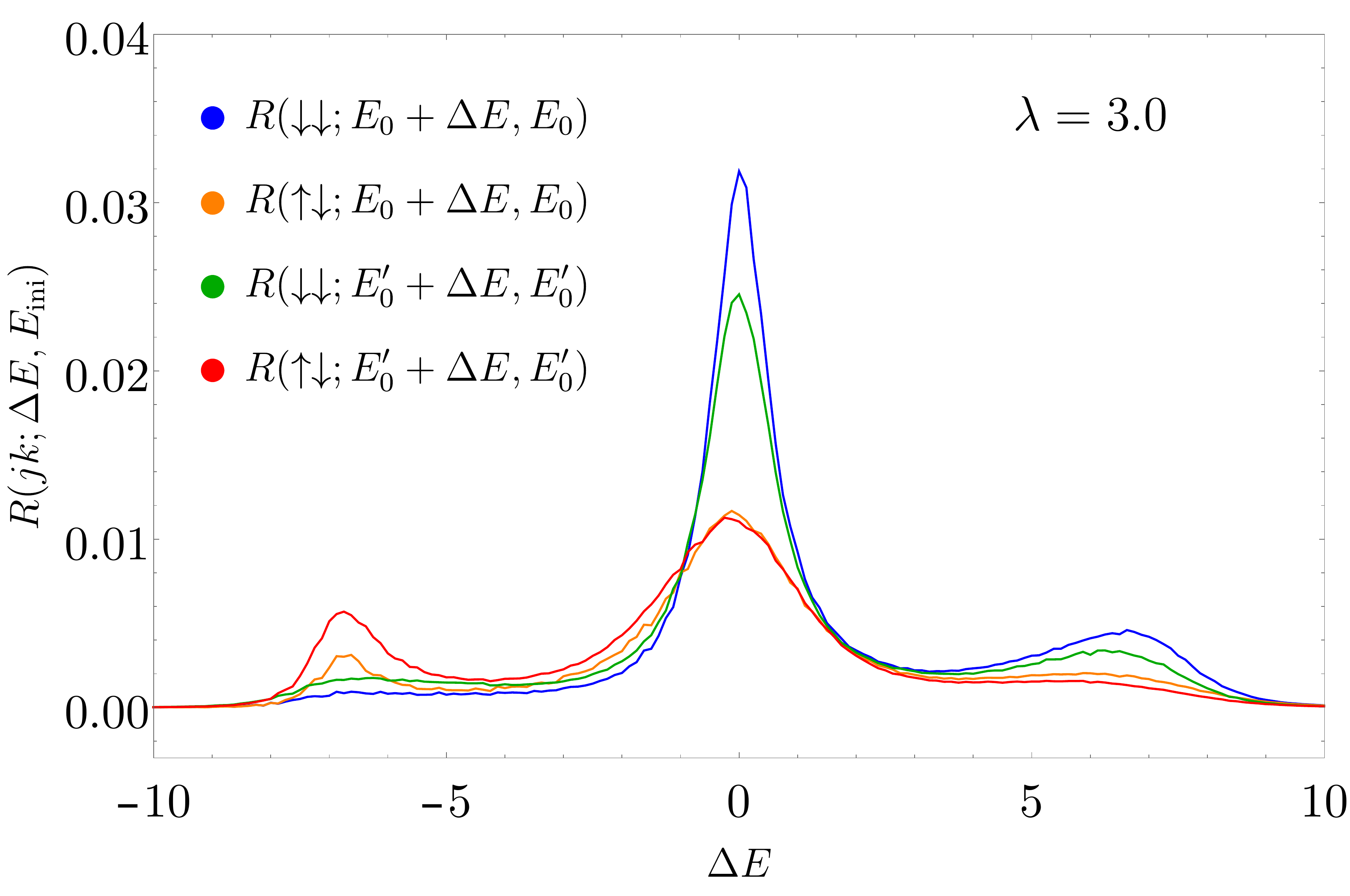}
  \caption{Transition probabilities for $\lambda=3.0$, $j=\,\uparrow,\downarrow$ and initial energy intervals $E_0, E_0'$. There are noticeable deviations between the transition probabilities for both initial energies, which indicates that stiffness is violated to some extent.}
  \label{stiff30}
\end{figure}

  \newpage
\noindent
The second setup under consideration is the transverse Ising model with defects. The system consists of a single site and the bath is a chain of length $L=14$, yielding a total of $15$ sites. The system, bath and interaction Hamiltonians are given by
\begin{equation}
H_\text{sys}=\omega\sigma^x_0\quad\quad H_\text{int} = \lambda\,\sigma^z_0\sigma^z_1
\end{equation}
\begin{equation*}
H_\text{bath}=g\sum_{i=1}^{L-1}\sigma^z_i\sigma^z_{i+1}+h\sum_{i=1}^{L}\sigma^x_i+\mu(h_{2}\sigma^z_2+h_{5}\sigma^z_5)\,,
\end{equation*}
where $\sigma^{x,y,z}_i$ denote Pauli matrices on site $i$. Again, due to symmetry, the eigenstates of the entropy production operator are equal to the eigenstates of the uncoupled Hamiltonian. The parameters are chosen as $
\omega=1.0$, $g=1.0$, $h=1.0$, $\mu=0.5$, $h_{2}=1.11$ and $h_{5}=1.61$, while the interaction strength $\lambda$ is varied. This choice of parameters ensures that the bath exhibits chaotic behavior. The initial bath energy is $\varepsilon_\text{bath}^{b_0}=-5.569$ corresponding to an inverse temperature $\beta=0.186$. The composite system is initialized in the product state
\begin{equation}
\label{ini2}
\rho(0) = |\hspace*{-2px}\downarrow,b_0\rangle\langle \downarrow,b_0|\,.
\end{equation}
The temporal behavior of the IFT quantity is shown in Fig. \ref{isingift} for various interaction strengths. For weak ($\lambda=0.3$) to moderate ($\lambda \leq 1.0$) interactions the IFT is practically fulfilled. For the strongest interaction $\lambda=3.0$ there exist small but noticeable deviations of about $0.05$ for longer times. Judging from Fig. \ref{isingift}, we would expect stiffness and smoothness to be sufficiently fulfilled for weak to moderate interaction strengths, while for the strongest interaction stiffness and/or smoothness should be violated to some extent.\\
\newpage
\noindent
Firstly, we investigate to what extent stiffness exists in the Ising model at hand by considering two particular interaction strengths, i.e. $\lambda=0.5$ and $\lambda=3.0$. For both cases we set $k=\,\downarrow$ and calculate transition probabilities for both $j=\,\downarrow,\uparrow$. We inspect two different initial energy intervals of the bath. The first interval $E_0$ is centered around $\varepsilon_\text{bath}^{b_0}$ with bin size $\Delta=0.125$. The second energy interval $E_0'$ is shifted by an amount $\Xi$ relative to $E_0$, i.e. $E_0' = E_0 + \Xi$. The shift $\Xi$ is equal to the arithmetic mean of standard deviations of the uncoupled Hamiltonian with respect to $n_b$ bath eigenstates, where $n_b$ is the number of eigenstates in the energy interval $E_0$.
\begin{equation}
\Xi = \dfrac{1}{n_b}\sum_b \sqrt{\langle \psi^{\downarrow,b}_t |H_\text{unc}^2|\psi^{\downarrow,b}_t \rangle - \langle\psi^{\downarrow,b}_t  |H_\text{unc}|\psi^{\downarrow,b}_t  \rangle^2}
\end{equation}
Here $\psi^{\downarrow,b}_t=\text{exp}(-\mathrm{i}Ht)|\hspace*{-5px}\downarrow,b\rangle$ is the final state of the composite system at the final time $t=100$.
For $\lambda=0.5$ ($\lambda=3.0$) we have $\Xi=0.702$ ($\Xi=3.343$). The combinations of $j=\,\uparrow, \downarrow$ and initial bath energies $E_0,E_0'$  yield four different curves for transition probabilities, which can be viewed in Fig. \ref{stiff5} for $\lambda=0.5$ and in Fig. \ref{stiff30} for $\lambda=3.0$. In the case of $\lambda=0.5$ the transition probabilities for $j=\,\uparrow$ as well as for $j=\,\downarrow$ exactly coincide for the different initial bath energies, indicating that stiffness exists for these combinations. In Fig. \ref{stiff30} the situation is different as there are clear discrepancies for both $j=\,\,\,\uparrow$ and $j=\,\downarrow$. This violation of stiffness may lead to the deviation for $\lambda=3.0$ in Fig. \ref{isingift}.\\

\noindent
Now we turn to the issue to what extent smoothness exists for the setup at hand. We tackle this question by means of a finite size scaling. 

\newpage
\noindent
We consider the averaged quantity in Eq. \eqref{ave}, where, again, we set $k=\,\downarrow$, the bin size to $\Delta=0.125$ and the initial bath energy interval to $E_0$ as above. We average over $j=\,\uparrow, \downarrow$ (thus, the $2$ in the denominator), over final energy intervals $A$ with $N_A=120$ and over $n_b=50$  bath eigenstates with energies neighboring $\varepsilon_\text{bath}^{b_0}$. \\

\noindent(For clarity, if convenient, we write the energy interval denoted by $E_\text{bath}^B$ as the argument of a transition probability $R$ instead of the capitalized index $B$. Additional commas and a semicolon are then used to improve readability.)
\begin{equation}
\label{ave}
\overline{\delta}= \sqrt{\dfrac{1}{2 N_A n_b}\sum_{j,A,b} |R(j\downarrow,Ab)-\overline{\vphantom{|}R}(j\downarrow;A,E_0)|^2}
\end{equation}

\begin{figure}[b]
  \includegraphics[width=\linewidth]{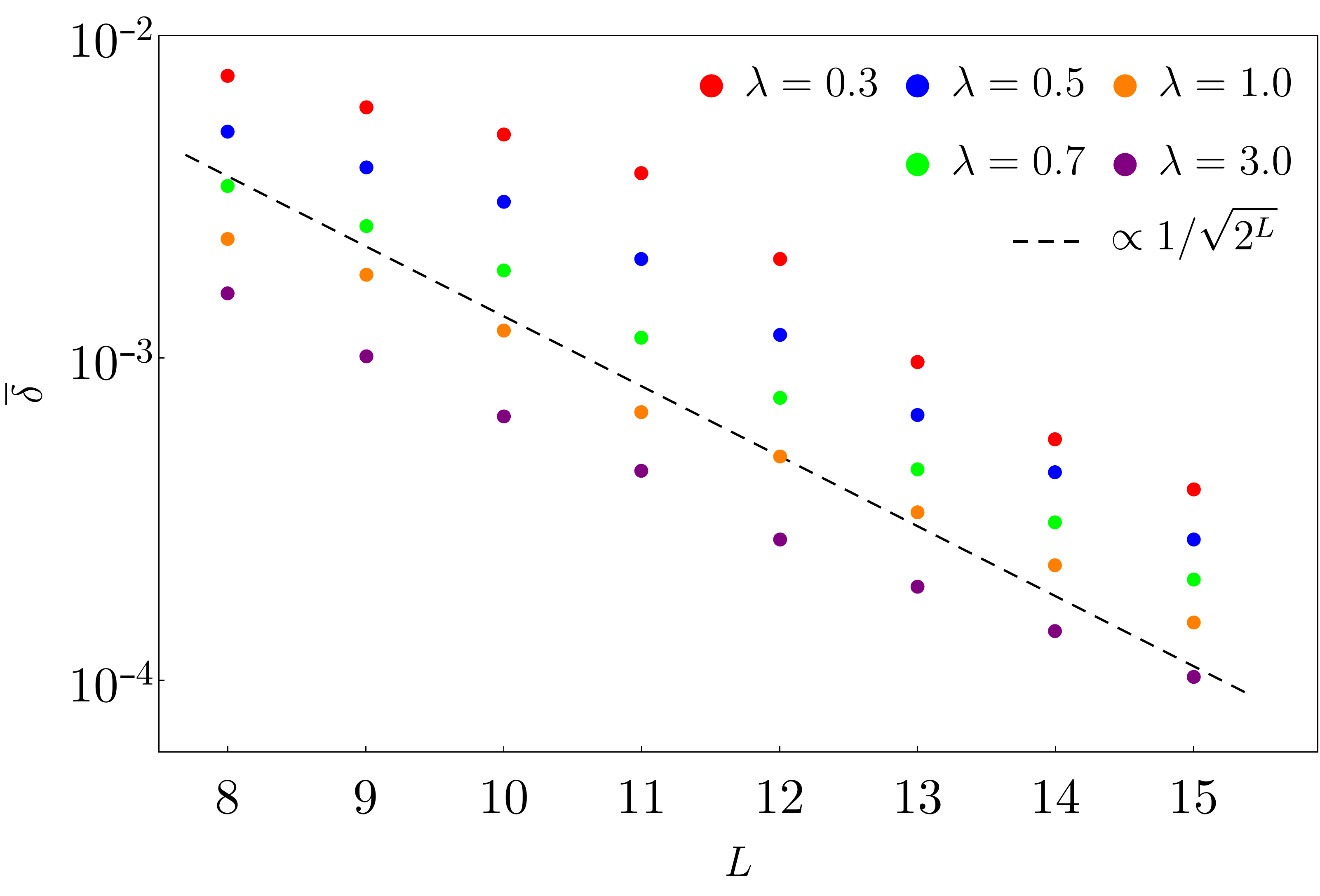}
  \caption{The averaged quantity $\overline{\delta}$ logarithmically plotted over the bath size $L$. The dashed black line serves as a guide to the eye. As the bath size increases, deviations of individual transition probabilities from the average decrease.}
  \label{smoooth}
\end{figure}

\newpage
\noindent
The result is depicted in Fig. \ref{smoooth}. Clearly, as the bath size increases, the transition probabilities become more and more equal to the respective average transition probability. The smoothness quantity $\overline{\delta}$ approximately scales as $\propto 1 / \sqrt{2^L}$. 
Note that $\overline{\delta}$ is smallest for $\lambda=3.0$. Thus, we conclude that the deviation of the IFT quantity from one for $\lambda=3.0$ in Fig. \ref{isingift} is probably due to the violation of stiffness, cf. Fig. \ref{stiff30}, and not due to a violation of smoothness.

\newpage
\section{Conclusion}
\label{conc}
\noindent
In this article we presented a way to derive the integral fluctuation theorem for microcanonical and pure quantum states under the assumption that the transition probabilities fulfill the properties of stiffness and smoothness. We numerically checked the validity of the IFT for two exemplary systems. Furthermore, the existence of stiffness and smoothness was directly scrutinized. This numerical analysis supports the idea that stiffness and smoothness are critical mechanisms enabling the validity of the IFT. In further work we plan to present a more thorough numerical analysis, checking stiffness and smoothness for a wide range of models and parameters.

\section*{Acknowledgments}
\label{conc}
\noindent
We thank T. Sagawa and E. Iyoda for extensive and fruitful discussions. As mentioned in the abstract, we recommend reading their submission to the arXiv about the origins of the IFT. This work was supported by the Deutsche  Forschungsgemeinschaft  (DFG)  within  the Research Unit FOR 2692 under Grant No. 397107022

\bibliography{literature}
\newpage

\end{document}